# Audacity of huge: overcoming challenges of data scarcity and data quality for machine learning in computational materials discovery

Aditya Nandy[1,2,§], Chenru Duan[1,2,§], and Heather J. Kulik[1,]*

[1]*Department of Chemical Engineering, Massachusetts Institute of Technology, Cambridge, MA 02139*
[1]*Department of Chemistry, Massachusetts Institute of Technology, Cambridge, MA 02139*
[§]*These authors contributed equally.*

**Corresponding Author:** email: hjkulik@mit.edu; mail: Department of Chemical Engineering, 77 Massachusetts Ave Rm 66-464, Cambridge, MA 02139, phone: 617-253-4584

[#]**email addresses:** nandy@mit.edu (A.N.); crduan@mit.edu (C.D.)

ABSTRACT: Machine learning (ML)-accelerated discovery requires large amounts of high-fidelity data to reveal predictive structure–property relationships. For many properties of interest in materials discovery, the challenging nature and high cost of data generation has resulted in a data landscape that is both scarcely populated and of dubious quality. Data-driven techniques starting to overcome these limitations include the use of consensus across functionals in density functional theory, the development of new functionals or accelerated electronic structure theories, and the detection of where computationally demanding methods are most necessary. When properties cannot be reliably simulated, large experimental data sets can be used to train ML models. In the absence of manual curation, increasingly sophisticated natural language processing and automated image analysis are making it possible to learn structure–property relationships from the literature. Models trained on these data sets will improve as they incorporate community feedback.

*Short title:* Addressing data limitations in materials discovery

**Introduction**

High-throughput computation or experiment coupled with machine learning (ML) has begun to address combinatorial challenges in materials discovery.[1-3] ML-accelerated discovery requires a large, high-fidelity data set. Data generation has benefited from recent advances in computing power and algorithms as well as the development of flow reactors, parallel experiments, and lab automation.[1,3] Computational (e.g., the Materials Project[4]) and experimental (e.g., Cambridge Structural Database (CSD)[5]) databases have made large data sets accessible for community use.[2]

Interest in materials discovery has moved toward harder targets and a focus on robust materials, where challenges arise for generation and curation strategies (Figure 1). Although density functional theory (DFT) is widely used for virtual high-throughput screening (VHTS), properties computed from DFT can be sensitive to the density functional approximation (DFA) used. DFA errors are often highest in promising functional materials classes that exhibit challenging electronic structure, instead requiring cost-prohibitive wavefunction theory (WFT) calculations.[6,7] Moreover, some properties of interest may be difficult to obtain from computation (e.g., synthesis outcomes or materials stability). High-throughput experimentation remains time-intensive relative to low-cost calculations and is often limited in scope to a single class of materials amenable to automated synthesis and characterization. With the exception of structural data, experimental properties are seldom reported by multiple sources in a standardized format.

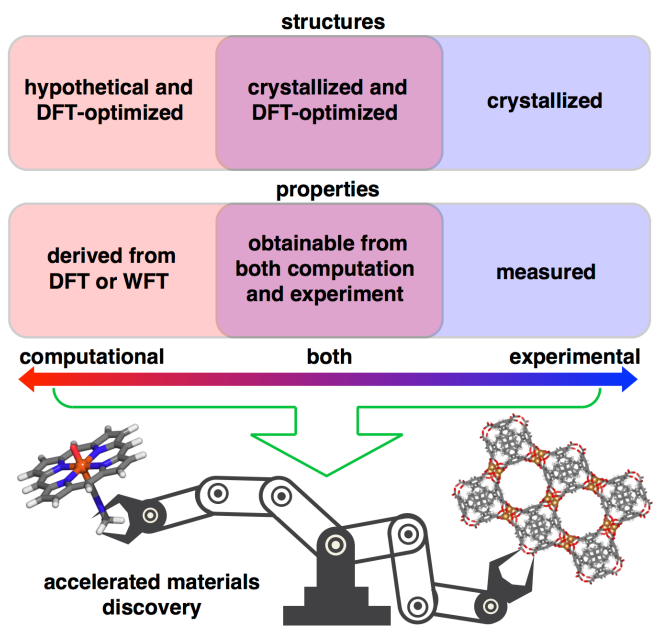

**Figure 1.** Overview of approaches to generate data for accelerated materials discovery. A depiction of the different data sources for structures and properties that can be derived from computation, experiments, or both to build large-scale databases for accelerated materials discovery.

In this Opinion, we describe how researchers are addressing challenges in data scarcity and data quality in ML-accelerated discovery: by leveraging data-driven methods for improving the fidelity of DFT properties, accelerating WFT calculations, and maximizing the utility of experimental data from multiple sources while obtaining community feedback.

**Addressing electronic structure method sensitivity**

Properties obtained with DFT depend on the choice of DFA, with no single DFA universally predictive for all materials.[7] DFAs are instead selected based on intuition or computational cost, thus introducing bias in data generation and reducing the quality of the data in a way that degrades utility for discovery efforts. To address this challenge, McAnanama-Brereton and Waller developed an approach[8] to identify optimal DFA-basis set combinations using game theory. They devised a three-player game addressing accuracy, complexity, and similarity of DFA-basis set combinations and solved the Nash equilibrium to yield the optimal combination. Gastegger *et al*.[9] applied a genetic algorithm (GA) to explore a space of popular DFAs and confirm that the GA is capable of identifying key components of a DFA that will be accurate on a benchmark dataset.

An alternative to choosing the functional most applicable to a given problem is to leverage consensus among predictions from multiple DFAs. Duan *et al*. computed[10] three properties for over 2,000 transition-metal complexes (TMCs) with 23 representative DFAs spanning multiple correlation families and "rungs" (e.g., semi-local to double hybrid). This study uncovered universal design rules that were invariant to DFA, basis, or data set choice from feature importance analysis of kernel ridge regression (KRR) models. They applied 23 ANNs, each fit to data from a single DFA, to discover spin-crossover (SCO) complexes, which have near-degenerate high-spin and low-spin states that are disproportionately sensitive to variations in DFA parameters.[11] By comparing hypothetical lead complexes to experimental SCO complexes, they noted that the leads recommended by a single-DFA-trained ANN (e.g., the B3LYP hybrid DFA) occupied a larger region of chemical space, indicating that many were false positives (Figure 2a). By requiring consensus among predictions of more than half of the DFA ANNs, they overcame this limitation of the single-DFA approach, producing robust (i.e., in agreement with experiment) candidate materials. However, a consensus-based approach may not be ideal for property prediction that benefits from error cancelation. Bartel *et al*.[12] compared seven ML models to predict formation energy and stability on 85,014 Materials Project Database compounds. They found that although these models have good accuracy on predicting formation energies comparable to a DFA, they lacked the error cancellation present in a standard DFA for predicting relative properties, suggesting caution in applying data-driven models blindly in materials discovery.

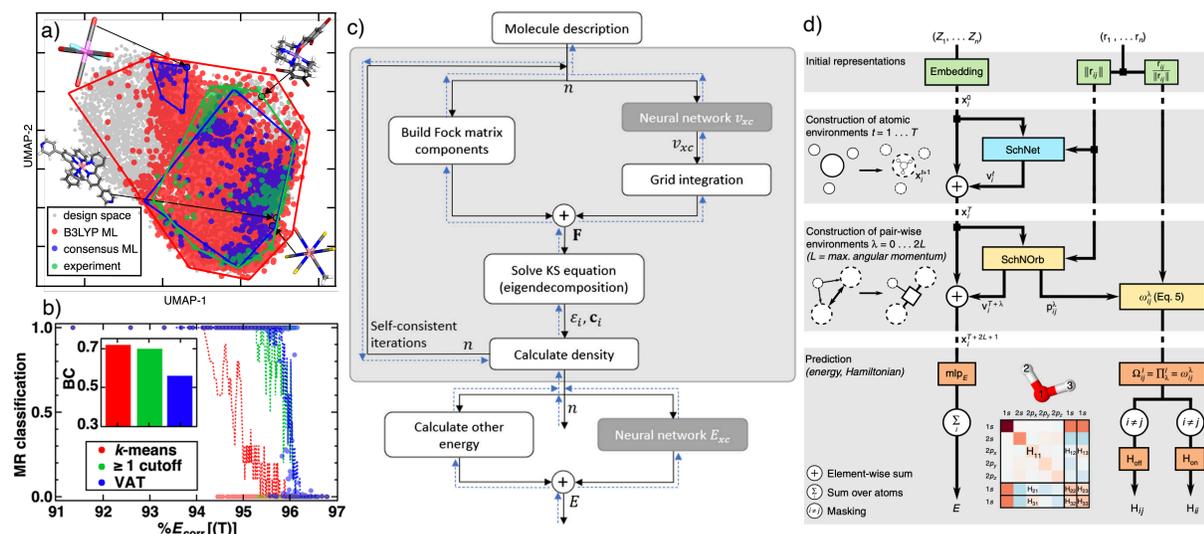

**Figure 2.** Machine learning approaches addressing fidelity limitations in quantum chemistry. a) uniform manifold approximation and projection (UMAP) visualization of SCO complexes from 187 200 TMCs reported by Duan *et al.*[10]: the entire design space (gray), leads predicted by a single NN trained on B3LYP data (red), by the consensus approach of NNs trained on 23 DFAs (blue), and experimental observation (green) with approximate convex hulls shown as solid lines. b) Performance of MR/SR classification on a set of 3165 equilibrium or distorted organic molecules using different methods reported by Duan *et al.*[13]: k-means clustering (red), a cutoff-based approach (green), and a semi-supervised learning method named virtual adversarial training (VAT, blue). c) Schematic of a fully differentiable KS-DFT framework reported by Kasim *et al.*,[14] with NNs representing a trainable exchange-correlation functional that yields both the electron density and energy. d) Illustration of the network architecture of SchNOrb developed by Schutt *et al.*[15], starting from initial representations of atom types and positions (top), continuing with the construction of representations of chemical environments of atoms and atom pairs (middle) before using these to predict energy and Hamiltonian matrix respectively (bottom). Reproduced with permission from studies reported by Duan *et al.* [10] and [13], Kasim *et al.*[14], and Schutt *et al.* [15], published by The Royal Society of Chemistry 2021, the American Chemical Society 2020, the American Physical Society 2021, and the Nature Publishing Group 2019, respectively.

Data-driven methods have augmented and supplanted the conventional approaches of trial and error or local fitting of parameters to revisit the search for a universal exchange-correlation functional.[16] Using Bayesian inference, a new DFA can be assembled as a linear combination of functional forms with statistically inferred coefficients, accelerating DFA design using known functional forms and alleviating the risk of overfitting during DFA parameterization.[17] ANNs have been used as an ansatz for designing DFAs due to their ability to represent any function. Brockherde *et al*. developed ML models that directly learn the ground-state density of a system, reducing the computational cost of solving the Kohn–Sham (KS) DFT equations iteratively.[18] By incorporating the KS equations as a regularization term in the loss function and providing feedback to ANNs in each training iteration, Nagai *et al*.[19] and Li *et al*.[20] demonstrated that the ANN functional can be learned with very few training molecules. Kasim *et al.* recast KS-DFT in a fully differentiable framework, enabling a density functional expressed as an ANN to be optimized with backpropagation, which demonstrated transferability to other small molecules containing elements and bond types not in the training set (Figure 2c).[14] These studies highlight the importance of imposing physical constraints during ML-based DFA design. Other emerging efforts in ML for DFT have extended to improving orbital-free DFT[21] and multiconfiguration pair-DFT.[22]

**Beyond DFT with ML**

Due to the use of a single Slater determinant of the non-interacting system in KS-DFT, DFT can fail to describe the electronic structure of systems that contain strong multireference (MR) character. Diagnostics to detect the degree of MR character using quantities of the wavefunction are often used to decide whether it is necessary to carry out a MR wavefunction theory (WFT) calculation. By investigating 3,000 equilibrium or distorted small organic molecules[23], Duan *et al*. found that different MR diagnostics seldom agree with each other (i.e., poor linear correlations), with high-cost WFT-based diagnostics better able to predict MR character than low-cost DFT diagnostics.[24] They observed consensus among MR diagnostics for the most extreme MR or single-reference (SR) points, motivating ML approaches that use only partially labeled data (i.e., semi-supervised learning).[13] A semi-supervised ANN classifier outperformed approaches typically used by experts to classify MR character (Figure 2b). Additionally, MR/SR classifiers were used to identify "DFT-safe" islands for chemical discovery, areas of chemical space where SR DFT predictions were expected to be of reasonable quality.[25]

In cases where strong MR character is detected, MR WFT methods are usually needed. These methods, however, are computationally demanding, despite some promising accelerations demonstrated by GPU-based parallel computing[26]. MR WFT also typically requires manual intervention. This challenge has started to be addressed by automated tools for active space selection using orbital entanglement analysis from loosely converged WFT calculations,[27] generalized valence bond orbitals,[28] a ranked orbital approach,[29] and ML models.[30]

ML has been used to accelerate WFT methods from an algorithmic perspective. For example, ANNs[31] and restricted Boltzmann machines[32] were applied to directly predict the coefficient of a configuration state function (CSF) in the iterative configuration interaction process to identify which CSFs could be pruned. Similarly, ANNs were used to initialize the guessed coupled cluster (CC) excitation amplitudes of a system using MP2-level molecular orbitals and one-electron integrals as inputs, reducing the number of iterations required.[33] In addition, ANNs[15] and KRR models[34] have been used to learn new representations of wavefunctions, enabling calculations of electron density, density of states, and dipole moments without explicit WFT calculations (Figure 2d). Hermann *et al*.[35] used ANNs as a wavefunction ansatz in quantum Monte Carlo, parameterizing a multi-determinant Slater-Jastrow-backflow type wavefunction. By applying the variational principle on this ANN wavefunction ansatz, they showed that the correlation energy can be quickly recovered (ca. 99%) for most of their test systems using very few (i.e., 10) determinants. To date most approaches have only been demonstrated on small systems with paired electrons, where MR WFT calculations can be performed easily. It is imperative to extend these developments to large systems with challenging electronic structure (e.g., metal–organic bonding) in order to impact ML-accelerated discovery of novel materials.

**ML with insight from experiments**

Many fundamental electronic properties, such as the ground-state spin of a TMC, remain challenging to determine by computation due to strong dependence on the method used. In some cases, a combination of experimental data and computation can overcome these limitations.[36,37] One of the largest sources of data is the CSD[5], which contains over 100,000 TMCs[11,38] and 90,000 metal–organic frameworks[39,40] (MOFs). Taylor *et al.* used an ANN[11] trained on DFT bond lengths to assign ground spin states to TMCs based on their CSD structures.[36] They confidently assigned spin states to around 90% of a large (ca. 2000) set of CSD Fe(II/III) TMCs by leveraging the relative DFA-invariance of DFT bond lengths in comparison to energetics.[36] This combined experiment-computation ML approach accelerated spin state assignment by orders of magnitude, avoiding decades of experimental effort (e.g., Mössbauer spectroscopy) that would otherwise be necessary.

Selectivity of a TMC catalyst is also difficult to predict solely by computation, because small barrier height differences (i.e., 1 kcal/mol) lead to divergent selectivity.[41] Santiago *et al.* built multiple linear regression (MLR) models combining descriptors from DFT calculations with experimental enantioselectivity data.[42] They demonstrated that DFT-derived physical organic descriptors[43] could predict experimental enantioselectivity. Maley *et al.* generalized this approach by using random forest[44] models trained on descriptors derived from DFT geometries to predict experimental selectivity and demonstrated the model for iterative ligand design.[45]

**ML for synthesis conditions**

The subtle and complex relationships that determine the optimal reaction conditions to achieve a desired chemical transformation are a poor fit for the predictive capabilities of low-cost first-principles computation. Intense effort has therefore focused on using ML to extract information on how conditions[46-48] such as temperature, time, and pH alter synthesis outcomes. The ChemDataExtractor toolkit[49,50] automates literature data extraction from thousands of manuscripts, including for use in generative models.[51] Kim *et al.* combined the ChemDataExtractor workflow with natural language processing (NLP) to identify how temperature and base concentration affect carbon nanotube formation.[52] For inorganic materials, Kim *et al.* and Jensen *et al.* have studied how precursor identities, precursor ratios, or additives[53] affect perovskite[48], oxide[52], and zeolite[54] formation.

**Insights from single-source experimental data**

Negative results are often underrepresented in the literature, and this positive publication bias creates a data imbalance in models trained on literature data. This has motivated single labs to carry out large experimental screens that generate both successful and failed experiments. Raccuglia *et al.* harnessed 3,955 completed reactions consisting of both successful and failed experiments and used this data on reaction conditions to inform future reactions.[55] Jia *et al.* curated a set of 548 experiments[56] on inorganic materials with randomly sampled synthesis conditions to demonstrate that the most popular synthesis conditions are not the most optimal.

Instead, optimizing synthesis conditions increased the surface area of the HKUST-1 MOF[57], consistent with observations that reported MOF surface areas generally increase as synthesis recipes are improved over time.[58] Demonstrating the benefits of curating a large data set under consistent conditions, Batra *et al.* investigated stability in the presence of water for 207 systematically synthesized MOFs and used ML to determine that metal ionization potential and ligand-to-metal ratio were predictive of MOF stability in water.[59] Yang *et al.* collected compositional and optical data on over 350,000 three-cation metal oxides to find materials that were stable in acidic conditions and active for electrocatalytic oxygen evolution, and they used statistical modeling to create maps between composition and optical properties.[60]

**Leveraging community data in ML**

When high-throughput, automated tools[57,60] are unavailable or incompatible with the quantity being curated, data collection can be limited in scope due to the effort required to perform each experiment, motivating instead a focus on community data resources like the CSD.[55,56,59] Taylor *et al.* curated[61] a set of bimetallic complexes from the CSD with emergent metal–metal interactions that are challenging to predict with first-principles DFT modeling. They used a subset of graph-based revised autocorrelation (RAC) descriptors to predict metal–metal bonding with KRR models, and they fit MLR models trained on RAC inputs to predict experimental redox potentials. Analysis of the most important features in the models revealed the overriding importance of metal group (i.e., electron configuration) rather than period in determining the properties of these complexes.

For MOFs, it is challenging to leverage all of the characterized structures in the CSD due to poor crystal structure quality and the presence of in-pore solvents. These limitations motivated the development of the Computation-Ready Experimental[62] (CoRE) MOF database, which contains sanitized experimental structures of nearly 12,000 MOFs (Figure 3). Moosavi *et al.* analyzed the chemical diversity of CoRE[62] MOFs and observed that experimental MOFs had significantly greater diversity in the metal secondary building unit (SBU) chemistry than was present in large (i.e., 100–300k) hypothetical MOF data sets[63] (Figure 3). They concluded that ML models built from limited hypothetical MOFs do not generalize for property prediction on diverse experimental structures. In another study, Jablonka *et al.* used user-assigned formal oxidation states of MOF SBU metals deposited in the CSD record to train a soft voting classifier ML model and identify where CSD oxidation states are likely incorrectly assigned[40].

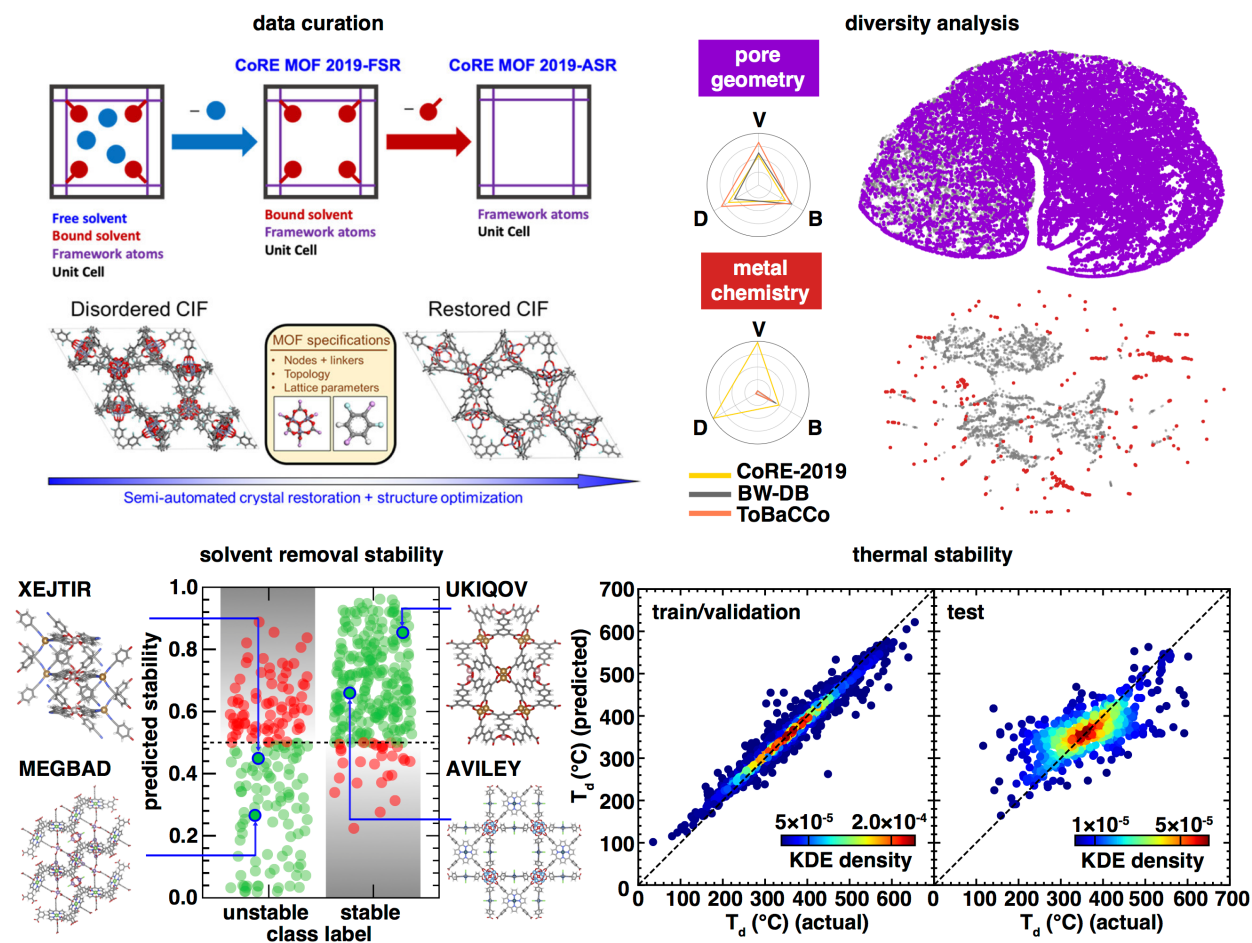

**Figure 3.** Data curation, analysis, and model prediction on MOFs. Data curation procedure for the CoRE MOF 2019[62] database. Experimental crystal structures were restored for subsequent

analysis (top left). Diversity analysis of hypothetical MOF spaces (colored in purple or red) relative to the full design space (gray) as t-SNE plots. Radar charts show three diversity metrics: variety (V), balance (B) and disparity (D), for three databases, CoRE MOF 2019[62], BW-DB[64], and ToBaCCo MOFs[65] (top right). Dot plots showing predicted activation stability (probability, no units from 0 for unstable to 1 for stable) vs actual class labels for MOFs in the test set for the solvent-removal stability data set. Data points are represented as translucent circles to depict data density and colored by the classification correctness: correct (green) and incorrect (red). Example structures and corresponding CSD refcodes for correct classifications are shown with blue outlines for two unstable MOFs: XEJTIR and MEGBAD, and two stable MOFs: UKIQOV and AVILEY (bottom left). Parity plots for predicting thermal decomposition temperatures in the thermal stability data set colored by kernel density estimation (KDE) density values, as indicated by inset color bars. In all cases, a black dashed parity line is shown (bottom right). Reproduced with permission from studies reported by Moosavi *et al.*[63]*,* Chung *et al.*[40], and Nandy *et al.*[66], published by the Nature Publishing Group 2020, the American Chemical Society 2019, the American Chemical Society 2021, respectively.

Nandy *et al.* expanded on this approach by using over 5,000 structures from the CoRE MOF database[62] and extracting data from the associated manuscripts rather than relying on information deposited as part of the CSD record.[66] They used NLP to determine material stability with respect to solvent removal (i.e., for activation) and trained ANN classifiers on this data set. They also automated the extraction of decomposition temperatures from thermogravimetric analysis (TGA) traces and trained an ANN regression model on these temperatures (Figure 3). At odds with existing heuristics[67], a MOF with a large pore volume was correctly predicted by the ML models to be stable.[66] Nandy *et al.* used the ML models to suggest alterations (e.g., linker fluorination or metal substitution) that should imbue stability in previously unstable MOFs.[66] NLP-based property extraction from the MOF literature is limited by the challenges of associating a measured property with a unique MOF name or structure.[68] Park *et al.* used heuristics[68] to identify MOF names without resorting to named-entity recognition.[69-72]

Although some properties present in the scientific literature can be extracted with NLP, spectra are reported in figures that cannot be parsed by NLP tools. Jiang *et al*. and Schwenker *et al*. have built ML models from hand-labeled data to identify subfigures within compound figures and classify their figure type (i.e. microscopy images, graphs, or illustrations).[73,74] Other work has automated identification of image length scales[74,75] to quantify particle distribution sizes or materials length scales.

**Community feedback on ML predictions**

Soliciting community feedback for ML models is essential for improving data fidelity and user confidence in model predictions, especially where subjectivity can be expected in the data. An effort to design organic light-emitting diodes used voting through a web interface to quantify synthetic accessibility of candidate materials.[76] The game theory density functional recommender by McAnanama-Brereton and Waller was incorporated into a web interface as a Turing test, collecting community feedback.[8] Similarly, Bennett *et al*. collected 12,553 data points on porous organic cage (POC) precursors labeled by three expert chemists to quantify synthesizability. They used this data to construct random forest ML classifiers to predict synthesizability of new POCs, replicating decisions made by expert synthetic chemists[77] (Figure 4).

**Figure 4.** Examples of community feedback interfaces. Sections of the MOFSimplify interface[78] for selecting a MOF for analysis and predicting properties using ANNs trained on experimental literature data (top left). Feedback interface of MOFSimplify for evaluating ANN model predictions (bottom left). Community survey questionnaire for quantifying MOF colors[79] (top right). Web interface for determining precursor synthesizability for POCs (bottom right). Reproduced with permission from studies reported by Nandy *et al.*[80], Jablonka *et al.*[81], and Bennett *et al.*[77], published by the Nature Publishing Group 2021, the Royal Society of Chemistry 2021, the American Chemical Society 2021, respectively.

Jablonka *et al.*[81] used a survey to quantify colors in CSD MOF descriptions. They gathered 4,184 quantitative RGB assignments for 162 qualitative colors used to describe MOF crystals (Figure 4). They analyzed this feedback and determined that colors such as beige corresponded to widely varying RGB values from different scientists, motivating a more quantitative scale for CSD color descriptions. Nandy *et al.* released a web interface[78] for improving the fidelity of NLP-derived data and testing user confidence in ML models that predict stability[66] of new MOFs.[80] To promote community-based active learning, the website solicits feedback on model predictions and encourages deposition of data (Figure 4). This feedback can improve the fidelity of NLP-extracted data and enrich data-poor regions of MOF chemical space.[80]

**Conclusion**

Although faster computational chemistry and robotic laboratory instrumentation have made it possible to obtain materials properties on a kilo-compound scale, the quest for novel and robust materials by ML-accelerated discovery poses new challenges for the scale and quality of data required from simulation and experiment. Recent efforts to address these limitations for high-fidelity VHTS have included using consensus among multiple DFAs,[10,11] applying ML to design new DFAs,[17,18] and accelerating WFT methods.[15,31] Researchers are using high-throughput experimentation and including failures to reduce bias when generating ML model training data,[57,60] parsing the literature with NLP tools to extract properties[52,53,66], and developing tools to automate analysis of graphical data.[73,74] As central tools for discovery of new materials, ML-accelerated workflows will deliver the greatest utility by soliciting and incorporating community feedback to enrich the underlying data and improve user confidence.[40,80]

DECLARATION OF INTEREST

The authors declare no conflict of interest.

CRediT authorship contribution statement

Chenru Duan: Visualization, Resources, Conceptualization, Writing – original draft, Writing – review & editing. Aditya Nandy: Visualization, Resources, Conceptualization,

Writing – original draft, Writing – review & editing. Heather J. Kulik: Conceptualization, Writing – original draft, Writing – review & editing.

ACKNOWLEDGMENT

This work was supported by the National Science Foundation grant numbers CBET-1704266 and CBET-1846426; the United States Department of Energy grant numbers DE-SC0012702, DE-SC0018096, DE-SC0019112, and DE-NA0003965; DARPA grant number D18AP00039; the Office of Naval Research grant numbers N00014-17-1-2956, N00014-18-1-2434, and N00014-20-1-2150; a National Science Foundation Graduate Research Fellowship under Grant #1122374 (to A.N.); an AAAS Marion Milligan Mason Award; and an Alfred P. Sloan Fellowship in Chemistry. The authors thank Adam H. Steeves and Vyshnavi Vennelakanti for providing a critical reading of the manuscript.

**References**

1. Dimitrov T, Kreisbeck C, Becker JS, Aspuru-Guzik A, Saikin SK: **Autonomous Molecular Design: Then and Now**. *ACS Appl. Mater. Interfaces* 2019, **11**:24825-24836
   http://dx.doi.org/10.1021/acsami.9b01226
2. Jablonka KM, Ongari D, Moosavi SM, Smit B: **Big-Data Science in Porous Materials: Materials Genomics and Machine Learning**. *Chem. Rev.* 2020, **120**:8066-8129
   http://dx.doi.org/10.1021/acs.chemrev.0c00004
3. Nandy A, Duan C, Taylor MG, Liu F, Steeves AH, Kulik HJ: **Computational Discovery of Transition-metal Complexes: From High-throughput Screening to Machine Learning**. *Chem. Rev.* 2021, **121**:9927-10000
   http://dx.doi.org/10.1021/acs.chemrev.1c00347
4. Jain A, Ong SP, Hautier G, Chen W, Richards WD, Dacek S, Cholia S, Gunter D, Skinner D, Ceder G, et al.: **Commentary: The Materials Project: A Materials Genome Approach to Accelerating Materials Innovation**. *APL Mater.* 2013, **1**:011002
   http://dx.doi.org/10.1063/1.4812323
5. Groom CR, Bruno IJ, Lightfoot MP, Ward SC: **The Cambridge Structural Database**. *Acta Crystallogr., Sect. B: Struct. Sci., Cryst. Eng. Mater.* 2016, **72**:171-179
   http://dx.doi.org/10.1107/s2052520616003954
6. Vogiatzis KD, Polynski MV, Kirkland JK, Townsend J, Hashemi A, Liu C, Pidko EA: **Computational Approach to Molecular Catalysis by 3d Transition Metals: Challenges and Opportunities**. *Chem. Rev.* 2019, **119**:2453-2523
   http://dx.doi.org/10.1021/acs.chemrev.8b00361

7. Janesko BG: **Replacing Hybrid Density Functional Theory: Motivation and Recent Advances**. *Chem. Soc. Rev.* 2021, **50**:8470-8495 http://dx.doi.org/10.1039/d0cs01074j
8. McAnanama-Brereton S, Waller MP: **Rational Density Functional Selection Using Game Theory**. *J. Chem. Inf. Model.* 2018, **58**:61-67 http://dx.doi.org/10.1021/acs.jcim.7b00542
9. Gastegger M, Gonzalez L, Marquetand P: **Exploring Density Functional Subspaces with Genetic Algorithms**. *Monatsh. Chem.* 2019, **150**:173-182 http://dx.doi.org/10.1007/s00706-018-2335-3
10. Duan C, Chen SX, Taylor MG, Liu F, Kulik HJ: **Machine Learning to Tame Divergent Density Functional Approximations: a New Path to Consensus Materials Design Principles**. *Chem. Sci.* 2021, **12**:13021-13036 http://dx.doi.org/10.1039/d1sc03701c

•• The authors computed three distinct properties for over 2000 transition metal complexes with 23 representative density functionals spanning multiple families and "rungs". Although the absolute properties differ by functionals, the linear correlations are strong, resulting universal design rules invariant to the choice of functional. Utilizing the consensus of multiple functionals, they uncovered spin-crossover complexes that are more in agreement with experimental observations compared to the conventional single-functional approach.

11. Janet JP, Liu F, Nandy A, Duan C, Yang TH, Lin S, Kulik HJ: **Designing in the Face of Uncertainty: Exploiting Electronic Structure and Machine Learning Models for Discovery in Inorganic Chemistry**. *Inorg. Chem.* 2019, **58**:10592-10606 http://dx.doi.org/10.1021/acs.inorgchem.9b00109
12. Bartel CJ, Trewartha A, Wang Q, Dunn A, Jain A, Ceder G: **A Critical Examination of Compound Stability Predictions from Machine-learned Formation Energies**. *npj Comput. Mater.* 2020, **6**:97 http://dx.doi.org/10.1038/s41524-020-00362-y
13. Duan C, Liu F, Nandy A, Kulik HJ: **Semi-supervised Machine Learning Enables the Robust Detection of Multireference Character at Low Cost**. *J. Phys. Chem. Lett.* 2020, **11**:6640-6648 http://dx.doi.org/10.1021/acs.jpclett.0c02018

• The authors demonstrated a semi-supervised learning method to distinguish systems that contain strong multi-reference character and thus density functional theory is likely to be inaccurate. Their model far outperforms the clustering methods and the cutoff-based methods that were widely used in the community. Combined with regression models that they previously built, their combined "decision engine" can make faithful classifications under only DFT costs.

14. Kasim MF, Vinko SM: **Learning the Exchange-Correlation Functional from Nature with Fully Differentiable Density Functional Theory**. *Phys. Rev. Lett.* 2021, **127**:126403 http://dx.doi.org/10.1103/PhysRevLett.127.126403

•• The authors developed a fully differentiable three-dimensional Kohn-Sham density functional theory framework in which the exchange-correlation functional is replaced by neural networks. They demonstrated that due to the fulfillment of Kohn-Sham equation between the learned density and energy, the model accuracy and transferability is greatly improved.

15. Schutt KT, Gastegger M, Tkatchenko A, Muller KR, Maurer RJ: **Unifying Machine Learning and Quantum Chemistry with a Deep Neural Network for Molecular Wavefunctions**. *Nat. Commun.* 2019, **10**:5024 http://dx.doi.org/10.1038/s41467-019-12875-2
16. Mardirossian N, Head-Gordon M: **Thirty Years of Density Functional Theory in Computational Chemistry: an Overview and Extensive Assessment of 200 Density Functionals**. *Mol. Phys.* 2017, **115**:2315-2372 http://dx.doi.org/10.1080/00268976.2017.1333644

17. Mitrofanov A, Korolev V, Andreadi N, Petrov V, Kalmykov S: **Simple Automatized Tool for Exchange-Correlation Functional Fitting**. *J. Phys. Chem. A* 2020, **124**:2700-2707 http://dx.doi.org/10.1021/acs.jpca.9b09093
18. Brockherde F, Vogt L, Li L, Tuckerman ME, Burke K, Muller KR: **Bypassing the Kohn-Sham Equations with Machine Learning**. *Nat. Commun.* 2017, **8**:872 http://dx.doi.org/10.1038/s41467-017-00839-3
19. Nagai R, Akashi R, Sugino O: **Completing Density Functional Theory by Machine Learning Hidden Messages from Molecules**. *npj Comput. Mater.* 2020, **6**:43 http://dx.doi.org/10.1038/s41524-020-0310-0
20. Li L, Hoyer S, Pederson R, Sun RX, Cubuk ED, Riley P, Burke K: **Kohn-Sham Equations as Regularizer: Building Prior Knowledge into Machine-Learned Physics**. *Phys. Rev. Lett.* 2021, **126**:036401 http://dx.doi.org/10.1103/PhysRevLett.126.036401
21. Meyer R, Weichselbaum M, Hauser AW: **Machine Learning Approaches toward Orbital-free Density Functional Theory: Simultaneous Training on the Kinetic Energy Density Functional and Its Functional Derivative**. *J. Chem. Theory Comput.* 2020, **16**:5685-5694 http://dx.doi.org/10.1021/acs.jctc.0c00580
22. King DS, Truhlar DG, Gagliardi L: **Machine-Learned Energy Functionals for Multiconfigurational Wave Functions**. *J. Phys. Chem. Lett.* 2021, **12**:7761-7767 http://dx.doi.org/10.1021/acs.jpclett.1c02042
23. Smith JS, Isayev O, Roitberg AE: **ANI-1, A Data Set of 20 Million Calculated Off-Equilibrium Conformations for Organic Molecules**. *Sci. Data* 2017, **4**:170193 http://dx.doi.org/10.1038/sdata.2017.193
24. Duan C, Liu F, Nandy A, Kulik HJ: **Data-Driven Approaches Can Overcome the Cost-Accuracy Trade-Off in Multireference Diagnostics**. *J. Chem. Theory Comput.* 2020, **16**:4373-4387 http://dx.doi.org/10.1021/acs.jctc.0c00358
25. Liu F, Duan C, Kulik HJ: **Rapid Detection of Strong Correlation with Machine Learning for Transition-Metal Complex High-Throughput Screening**. *J. Phys. Chem. Lett.* 2020, **11**:8067-8076 http://dx.doi.org/10.1021/acs.jpclett.0c02288
26. Fales BS, Martinez TJ: **Efficient Treatment of Large Active Spaces through Multi-GPU Parallel Implementation of Direct Configuration Interaction**. *J. Chem. Theory Comput.* 2020, **16**:1586-1596 http://dx.doi.org/10.1021/acs.jctc.9b01165
• The authors have extended the graphical processing unit-accelerated direct configuration interaction program to multiple devices, hugely reducing time spent on configuration interaction calculations and one- and two-particle reduced density matrix formation. They demonstrated that the iteration time of a configuration space of 165 million determinants is reduced only 3 seconds using NVIDIA P100 GPUs.
27. Stein CJ, Reiher M: **Automated Selection of Active Orbital Spaces**. *J. Chem. Theory Comput.* 2016, **12**:1760-1771 http://dx.doi.org/10.1021/acs.jctc.6b00156
28. Zou JX, Niu K, Ma HB, Li SH, Fang WH: **Automatic Selection of Active Orbitals from Generalized Valence Bond Orbitals**. *J. Phys. Chem. A* 2020, **124**:8321-8329 http://dx.doi.org/10.1021/acs.jpca.0c05216
29. King DS, Gagliardi L: **A Ranked-Orbital Approach to Select Active Spaces for High-Throughput Multireference Computation**. *J. Chem. Theory Comput.* 2021, **17**:2817-2831 http://dx.doi.org/10.1021/acs.jctc.1c00037
30. Jeong W, Stoneburner SJ, King D, Li RY, Walker A, Lindh R, Gagliardi L: **Automation of Active Space Selection for Multireference Methods via Machine Learning on **Chemical Bond Dissociation**. *J. Chem. Theory Comput.* 2020, **16**:2389-2399 http://dx.doi.org/10.1021/acs.jctc.9b01297
31. Coe JP: **Machine Learning Configuration Interaction**. *J. Chem. Theory Comput.* 2018, **14**:5739-5749 http://dx.doi.org/10.1021/acs.jctc.8b00849
32. Yang PJ, Sugiyama M, Tsuda K, Yanai T: **Artificial Neural Networks Applied as Molecular Wave Function Solvers**. *J. Chem. Theory Comput.* 2020, **16**:3513-3529 http://dx.doi.org/10.1021/acs.jctc.9b01132
33. Townsend J, Vogiatzis KD: **Data-Driven Acceleration of the Coupled-Cluster Singles and Doubles Iterative Solver**. *J. Phys. Chem. Lett.* 2019, **10**:4129-4135 http://dx.doi.org/10.1021/acs.jpclett.9b01442
34. Peyton BG, Briggs C, D'Cunha R, Margraf JT, Crawford TD: **Machine-Learning Coupled Cluster Properties through a Density Tensor Representation**. *J. Phys. Chem. A* 2020, **124**:4861-4871 http://dx.doi.org/10.1021/acs.jpca.0c02804
35. Hermann J, Schatzle Z, Noe F: **Deep-Neural-Network Solution of the Electronic Schrodinger Equation**. *Nat. Chem.* 2020, **12**:891-897 http://dx.doi.org/10.1038/s41557-020-0544-y
• The authors used neural networks as a wavefunction ansatz in quantum Monte Carlo calculations, where they parameterized a multi-determinant Slater-Jastrow-backflow type wavefunction using an NN. They then applied variational principle on this NN wavefunction ansatz and showed that the correlation energy can be quickly recovered up to 99% with very few determinants for most of their test systems.
36. Taylor MG, Yang T, Lin S, Nandy A, Janet JP, Duan C, Kulik HJ: **Seeing Is Believing: Experimental Spin States from Machine Learning Model Structure Predictions**. *J. Phys. Chem. A* 2020, **124**:3286-3299 http://dx.doi.org/10.1021/acs.jpca.0c01458
37. Rosen AS, Iyer SM, Ray D, Yao Z, Aspuru-Guzik A, Gagliardi L, Notestein JM, Snurr RQ: **Machine Learning the Quantum-Chemical Properties of Metal–Organic Frameworks for Accelerated Materials Discovery**. *Matter* 2021, **4**:1578-1597 http://dx.doi.org/10.1016/j.matt.2021.02.015
38. Balcells D, Skjelstad BB: **tmQM Dataset—Quantum Geometries and Properties of 86k Transition Metal Complexes**. *J. Chem. Inf. Model.* 2020, **60**:6135-6146 http://dx.doi.org/10.1021/acs.jcim.0c01041
39. Sarkisov L, Bueno-Perez R, Sutharson M, Fairen-Jimenez D: **Materials Informatics with PoreBlazer v4.0 and the CSD MOF Database**. *Chem. Mater.* 2020, **32**:9849-9867 http://dx.doi.org/10.1021/acs.chemmater.0c03575
40. Jablonka KM, Ongari D, Moosavi SM, Smit B: **Using Collective Knowledge to Assign Oxidation States of Metal Cations in Metal–Organic Frameworks**. *Nat. Chem.* 2021, **13**:771-777 http://dx.doi.org/10.1038/s41557-021-00717-y
41. Zahrt AF, Henle JJ, Rose BT, Wang Y, Darrow WT, Denmark SE: **Prediction of Higher-Selectivity Catalysts by Computer-Driven Workflow and Machine Learning**. *Science* 2019, **363**:1-11 http://dx.doi.org/10.1126/science.aau5631
42. Santiago CB, Guo J-Y, Sigman MS: **Predictive and Mechanistic Multivariate Linear Regression Models for Reaction Development**. *Chem. Sci.* 2018, **9**:2398-2412 http://dx.doi.org/10.1039/c7sc04679k
43. Durand DJ, Fey N: **Computational Ligand Descriptors for Catalyst Design**. *Chem. Rev.* 2019, **119**:6561-6594 http://dx.doi.org/10.1021/acs.chemrev.8b00588
44. Breiman L: **Random Forests**. *Machine Learning* 2001, **45**:5−32

45. Maley Steven M, Kwon D-H, Rollins N, Stanley JC, Sydora OL, Bischof SM, Ess DH: **Quantum-Mechanical Transition-State Model Combined with Machine Learning Provides Catalyst Design Features for Selective Cr Olefin Oligomerization**. *Chem. Sci.* 2020, **11**:9665-9674 http://dx.doi.org/10.1039/d0sc03552a
46. Tshitoyan V, Dagdelen J, Weston L, Dunn A, Rong Z, Kononova O, Persson KA, Ceder G, Jain A: **Unsupervised Word Embeddings Capture Latent Knowledge from Materials Science Literature**. *Nature* 2019, **571**:95-98 http://dx.doi.org/10.1038/s41586-019-1335-8
47. Park H, Kang Y, Choe W, Kim J: **Mining Insights on Metal-Organic Framework Synthesis from Scientific Literature Texts**. *arXiv* 2021, http://dx.doi.org/arXiv:2108.13590
48. Kim E, Jensen Z, van Grootel A, Huang K, Staib M, Mysore S, Chang H-S, Strubell E, McCallum A, Jegelka S, et al.: **Inorganic Materials Synthesis Planning with Literature-Trained Neural Networks**. *J. Chem. Inf. Model.* 2020, **60**:1194-1201 http://dx.doi.org/10.1021/acs.jcim.9b00995
49. Swain MC, Cole JM: **ChemDataExtractor: A Toolkit for Automated Extraction of Chemical Information from the Scientific Literature**. *J. Chem. Inf. Model.* 2016, **56**:1894-1904 http://dx.doi.org/10.1021/acs.jcim.6b00207
50. Mavračić J, Court CJ, Isazawa T, Elliott SR, Cole JM: **ChemDataExtractor 2.0: Autopopulated Ontologies for Materials Science**. *J. Chem. Inf. Model.* 2021, **61**:4280-4289 http://dx.doi.org/10.1021/acs.jcim.1c00446
51. Court CJ, Jain A, Cole JM: **Inverse Design of Materials That Exhibit the Magnetocaloric Effect by Text-Mining of the Scientific Literature and Generative Deep Learning**. *Chem. Mater.* 2021, http://dx.doi.org/10.1021/acs.chemmater.1c01368
52. Kim E, Huang K, Saunders A, McCallum A, Ceder G, Olivetti E: **Materials Synthesis Insights from Scientific Literature via Text Extraction and Machine Learning**. *Chem. Mater.* 2017, **29**:9436-9444 http://dx.doi.org/10.1021/acs.chemmater.7b03500
53. Jensen Z, Kwon S, Schwalbe-Koda D, Paris C, Gómez-Bombarelli R, Román-Leshkov Y, Corma A, Moliner M, Olivetti EA: **Discovering Relationships between OSDAs and Zeolites through Data Mining and Generative Neural Networks**. *ACS Cent. Sci.* 2021, **7**:858-867 http://dx.doi.org/10.1021/acscentsci.1c00024

••The authors mine the identities of organic structure directing agents (OSDAs) from the literature. They use the mined data to draw correlations between OSDAs and final zeolite topologies. They then used this knowledge and a generative model to suggest new OSDAs that could be used to aid zeolite synthesis.

54. Jensen Z, Kim E, Kwon S, Gani TZH, Román-Leshkov Y, Moliner M, Corma A, Olivetti E: **A Machine Learning Approach to Zeolite Synthesis Enabled by Automatic Literature Data Extraction**. *ACS Cent. Sci.* 2019, **5**:892-899 http://dx.doi.org/10.1021/acscentsci.9b00193
55. Raccuglia P, Elbert KC, Adler PDF, Falk C, Wenny MB, Mollo A, Zeller M, Friedler SA, Schrier J, Norquist AJ: **Machine-Learning-Assisted Materials Discovery Using Failed Experiments**. *Nature* 2016, **533**:73-76 http://dx.doi.org/10.1038/nature17439
56. Jia X, Lynch A, Huang Y, Danielson M, Lang'at I, Milder A, Ruby AE, Wang H, Friedler SA, Norquist AJ, et al.: **Anthropogenic Biases in Chemical Reaction Data Hinder Exploratory Inorganic Synthesis**. *Nature* 2019, **573**:251-255 http://dx.doi.org/10.1038/s41586-019-1540-5

• The authors used single-source data to conclude that the most popular synthesis conditions are not the most optimal and are oversampled due to bias from publications. Instead, the authors found better conditions with a random search of the synthesis parameter space.

57. Moosavi SM, Chidambaram A, Talirz L, Haranczyk M, Stylianou KC, Smit B: **Capturing Chemical Intuition in Synthesis of Metal-Organic Frameworks**. *Nat. Commun.* 2019, **10** http://dx.doi.org/10.1038/s41467-019-08483-9
58. Agrawal M, Han R, Herath D, Sholl DS: **Does Repeat Synthesis in Materials Chemistry Obey a Power Law?** *Proc. Natl. Acad. Sci. U. S. A.* 2020, **117**:877-882 http://dx.doi.org/10.1073/pnas.1918484117
59. Batra R, Chen C, Evans TG, Walton KS, Ramprasad R: **Prediction of Water Stability of Metal–Organic Frameworks Using Machine Learning**. *Nat. Mach. Intell.* 2020, **2**:704-710 http://dx.doi.org/10.1038/s42256-020-00249-z
60. Yang L, Haber JA, Armstrong Z, Yang SJ, Kan K, Zhou L, Richter MH, Roat C, Wagner N, Coram M, et al.: **Discovery of Complex Oxides via Automated Experiments and Data Science**. *Proc. Natl. Acad. Sci. U. S. A.* 2021, **118**:e2106042118 http://dx.doi.org/10.1073/pnas.2106042118
61. Taylor MG, Nandy A, Lu CC, Kulik HJ: **Deciphering Cryptic Behavior in Bimetallic Transition Metal Complexes with Machine Learning**. *J. Phys. Chem. Lett.* 2021, **12**:9812-9820 http://dx.doi.org/arXiv:2107.14280
62. Chung YG, Haldoupis E, Bucior BJ, Haranczyk M, Lee S, Zhang H, Vogiatzis KD, Milisavljevic M, Ling S, Camp JS, et al.: **Advances, Updates, and Analytics for the Computation-Ready, Experimental Metal–Organic Framework Database: CoRE MOF 2019**. *J. Chem. Eng. Data* 2019, **64**:5985-5998 http://dx.doi.org/10.1021/acs.jced.9b00835
63. Moosavi SM, Nandy A, Jablonka KM, Ongari D, Janet JP, Boyd PG, Lee Y, Smit B, Kulik HJ: **Understanding the Diversity of the Metal-Organic Framework Ecosystem**. *Nat. Commun.* 2020, **11** http://dx.doi.org/10.1038/s41467-020-17755-8
64. Boyd PG, Chidambaram A, García-Díez E, Ireland CP, Daff TD, Bounds R, Gładysiak A, Schouwink P, Moosavi SM, Maroto-Valer MM, et al.: **Data-Driven Design of Metal–Organic Frameworks for Wet Flue Gas $CO_2$ Capture**. *Nature* 2019, **576**:253-256 http://dx.doi.org/10.1038/s41586-019-1798-7
65. Colón YJ, Gómez-Gualdrón DA, Snurr RQ: **Topologically Guided, Automated Construction of Metal–Organic Frameworks and Their Evaluation for Energy-Related Applications**. *Cryst. Growth Des.* 2017, **17**:5801-5810 http://dx.doi.org/10.1021/acs.cgd.7b00848
66. Nandy A, Duan C, Kulik HJ: **Using Machine Learning and Data Mining to Leverage Community Knowledge for the Engineering of Stable Metal-Organic Frameworks**. *J. Am. Chem. Soc.* 2021, **143**:17535-17547

•• The authors used natural language processing to gather experimental stability data and trained predictive machine learning models to identify when a metal-organic framework (MOF) is thermally stable and stable upon activation. They demonstrated that their models can be used on new experimentally characterized MOFs.

67. Ayoub G, Islamoglu T, Goswami S, Friščić T, Farha OK: **Torsion Angle Effect on the Activation of UiO Metal–Organic Frameworks**. *ACS Appl. Mater. Interfaces* 2019, **11**:15788-15794 http://dx.doi.org/10.1021/acsami.9b02764

68. Park S, Kim B, Choi S, Boyd PG, Smit B, Kim J: **Text Mining Metal–Organic Framework Papers**. *J. Chem. Inf. Model.* 2018, **58**:244-251 http://dx.doi.org/10.1021/acs.jcim.7b00608
69. Bucior BJ, Rosen AS, Haranczyk M, Yao Z, Ziebel ME, Farha OK, Hupp JT, Siepmann JI, Aspuru-Guzik A, Snurr RQ: **Identification Schemes for Metal–Organic Frameworks To Enable Rapid Search and Cheminformatics Analysis**. *Cryst. Growth Des.* 2019, **19**:6682-6697 http://dx.doi.org/10.1021/acs.cgd.9b01050
70. Kononova O, He T, Huo H, Trewartha A, Olivetti EA, Ceder G: **Opportunities and Challenges of Text Mining in Materials Research**. *iScience* 2021, **24**:102155 http://dx.doi.org/10.1016/j.isci.2021.102155
71. Weston L, Tshitoyan V, Dagdelen J, Kononova O, Trewartha A, Persson KA, Ceder G, Jain A: **Named Entity Recognition and Normalization Applied to Large-Scale Information Extraction from the Materials Science Literature**. *J. Chem. Inf. Model.* 2019, **59**:3692-3702 http://dx.doi.org/10.1021/acs.jcim.9b00470
72. Cole JM: **A Design-to-Device Pipeline for Data-Driven Materials Discovery**. *Acc. Chem. Res.* 2020, **53**:599-610 http://dx.doi.org/10.1021/acs.accounts.9b00470
73. Jiang W, Schwenker E, Spreadbury T, Ferrier N, Chan MKY, Cossairt O: **A Two-stage Framework for Compound Figure Separation**. *arXiv* 2021, http://dx.doi.org/arXiv:2101.09903
74. Schwenker E, Jiang W, Spreadbury T, Ferrier N, Cossairt O, Chan MKY: **EXSCLAIM! -- An Automated Pipeline for the Construction of Labeled Materials Imaging Datasets from Literature**. *arXiv* 2021, http://dx.doi.org/arXiv:2103.10631

•• The authors describe methods for deconstructing a complex figure into its corresponding subfigures. They use machine learning models trained with hand-labeled data to categorize the subfigures and use natural language processing to capture and store information from figure captions.

75. Mukaddem KT, Beard EJ, Yildirim B, Cole JM: **ImageDataExtractor: A Tool To Extract and Quantify Data from Microscopy Images**. *J. Chem. Inf. Model.* 2019, **60**:2492-2509 http://dx.doi.org/10.1021/acs.jcim.9b00734
76. Gómez-Bombarelli R, Aguilera-Iparraguirre J, Hirzel TD, Duvenaud D, Maclaurin D, Blood-Forsythe MA, Chae HS, Einzinger M, Ha D-G, Wu T, et al.: **Design of Efficient Molecular Organic Light-Emitting Diodes by a High-Throughput Virtual Screening and Experimental Approach**. *Nat. Mater.* 2016, **15**:1120-1127 http://dx.doi.org/10.1038/nmat4717
77. Bennett S, Szczypiński F, Turcani L, Briggs M, Greenaway RL, Jelfs K: **Materials Precursor Score: Modeling Chemists' Intuition for the Synthetic Accessibility of Porous Organic Cage Precursors**. *J. Chem. Inf. Model.* 2021, **61**:4342-4356 http://dx.doi.org/10.1021/acs.jcim.1c00375

• The authors solicited feedback from expert synthetic organic chemists on the synthesizability of POCs from precursors. They were able to train models and assign a materials precursor score (MPScore) that could identify POCs that contained easy-to-synthesize building blocks and rule out POCs that contained difficult-to-synthesize building blocks.

78. Nandy A, Terrones G, Arunachalam N, Duan C, Kastner DW, Kulik HJ: **MOFSimplify**. 2021, https://mofsimplify.mit.edu
79. Jablonka KM, Moosavi SM, Asgari M, Ireland C, Patiny L, Smit B: **Color Jeopardy**. 2021, https://colorjeopardy.herokuapp.com/

80. Nandy A, Terrones G, Arunachalam N, Duan C, Kastner DW, Kulik HJ: **MOFSimplify: Machine Learning Models with Extracted Stability Data of Three Thousand Metal-Organic Frameworks**. *arXiv* 2021, http://dx.doi.org/arXiv:2109.08098

81. Jablonka KM, Moosavi SM, Asgari M, Ireland C, Patiny L, Smit B: **A Data-Driven Perspective on the Colours of Metal–Organic Frameworks**. *Chem. Sci.* 2021, **12**:3587-3598 http://dx.doi.org/10.1039/d0sc05337f

• The authors of this manuscript use decades of structural data and their corresponding labeled metal cation oxidation states to train models that relate chemistry to oxidation state. They use these models to detect when the oxidation states are inconsistent with other materials that have similar metal-local environments. The authors use this method to determine incorrectly assigned oxidation states of metal-organic frameworks.